\documentclass[%
 aip,
 numerical,
 jcp,
 floatfix,
rsi,%
amsmath,amssymb,
reprint,%
author-year%
]{revtex4-1}
\usepackage{xcolor}
\usepackage{graphicx}
\usepackage{dcolumn}
\usepackage{bm}
\usepackage{natbib}
\usepackage{hyperref}

\bibpunct{(}{)}{,}{n}{,}{,}

\begin{document}

\title{Phase-space propagation and stability analysis of the 1-dimensional Schr\"odinger equation for finding bound and resonance states of rotationally excited H$_2$.}

\author{Juan S. Molano}
\affiliation{Department of Chemical Sciences, Universidad Icesi, Cali, Colombia}

\author{Carlos A. Arango}
\email{caarango@icesi.edu.co}
\affiliation{Department of Chemical Sciences, Universidad Icesi, Cali, Colombia}

\date{\today}

\begin{abstract}
A phase-space representation of the 1-dimensional Schr\"odinger equation is employed to obtain bound and resonance states of rotationally excited  H$_2$. The structure of the phase-space tangent field is analyzed and related to the behavior of the wave function in classically allowed and forbidden regions. In this phase-space representation, bound states behave like unstable orbits meanwhile resonance states behave like asymptotically stable cycles. The arc length and winding number of the phase-space trajectories, as functions of the energy, are used to obtain the energy eigenvalues of bound and resonance states of H$_2$.

\end{abstract}


\maketitle 

\section{Introduction}

The 1-dimensional time independent nonrelativistic Schr\"odinger equation (TISE) is a real second order linear differential equation in the position basis [\onlinecite{Schrodinger1926}]. As a second order differential equation, two boundary conditions (BCs) must be specified in order to obtain an unique solution of the TISE [\onlinecite{Derrick1997}]. Boundary conditions must fit the physical problem of interest, for example, Dirichlet BCs are suitable to  obtain bound states meanwhile Cauchy BCs are better to obtain continuum or resonance states. 

Analytical solutions of the TISE are possible only for few model systems. The rigid rotor, the harmonic oscillator, the hydrogen atom, and the Morse oscillator are some of the exactly solvable models of interest in Chemical Physics [\onlinecite{Atkins2011}]. In general, it is impossible to find explicit solutions of the TISE, and numerical methods are needed. Finite differences [\onlinecite{LeVeque2007,Hairer2009,Hairer1996}], spectral, and pseudospectral numerical methods are broadly employed to solve Schr\"odinger's equation [\onlinecite{Tannor2007}]. Pseudospectral methods employ a basis of spatially localized functions centered at different spatial grid points.  Bound and resonance states of the TISE are efficiently obtained by pseudospectral methods like Discrete Value Representation (DVR) [\onlinecite{Light1985,Colbert1992,Molano2019}] and Fourier Grid Hamiltonian (FGH) [\onlinecite{Marston1989,Chu1990,Yao1993}]. These pseudospectral methods are better suitable for Dirichlet boundary conditions, and need of numerical strategies, as complex scaling or complex absorbing potential (CAP), to obtain resonance or continuum states [\onlinecite{Bardsley1978,Moiseyev1979,Muga2004,Landau2015}]. Spectral and pseudospectral methods for solving the TISE on unbound domains, infinite or semi-infinite, are beginning to show interesting and promising results in spite of their complexity [\onlinecite{Boyd1990,Boyd2000,Alici2015,Alici2020}]. Numerical solutions of the TISE can be obtained also by Fourier analysis of time correlation functions produced by propagating a suitable initial wavepacket with the Time Dependent Schr\"odinger Equation (TDSE). This time dependent approach has been used to calculate bound and resonance states in atomic and molecular systems \cite{Blinder1964,Kosloff1994}. The TDSE equation is a partial differential equation (PDE), and although the time propagation is an initial value problem, the spatial part of the PDE requires of well defined Dirichlet BCs for bound states, and optical potentials for resonances or continuum states \cite{Montano2000,Zambrano2002}.

As a second order ordinary differential equation (ODE), the 1-dimensional TISE can be represented in phase space [\onlinecite{Arnold1992}].  In this representation a second order ODE is replaced by a set of two first order ODEs with Cauchy BCs instead of the Dirichlet BCs of the original problem. Numerically this idea is implemented in Numerov's method to obtain both bound and resonance states [\onlinecite{Numerov1927,Kobeissi1988,Simos1997,Tselyaev2004,VigoAguiar2005,Pillai2012}]. 

In this paper we study the  H$_2$ molecule, with rotational excitation, employing the phase-space representation of the 1-dimensional TISE. We analyze the relationship between the structure of the phase-space vector field and the onset of bound and resonance states. Bound and resonance states are obtained as trajectories in phase-space. We show that eigenstates of the TISE can be obtained by analyzing the winding number and the arc-length of the trajectories as functions of the energy. 

\section{Theory and Methods}
The radial Schr\"odinger equation for a diatomic molecule with reduced mass $\mu$, internuclear distance $r\in(0,\infty)$, and rotational angular momentum $l\in\mathbb{Z}$, $l\ge0$, is given by 
\begin{equation}\label{Schrodinger}
    \left(-\frac{1}{2\mu}\frac{d^2}{dr^2}+V_l(r)\right)\psi(r)=\mathcal{E}\psi(r).
\end{equation}
As a singular Sturm-Liouville problem, the energy spectrum of \eqref{Schrodinger} consists of a countable set of eigenvalues and eigenfunctions, $\mathcal{E}_n$ and $\psi_n$, and a continuum [\onlinecite{Kodaira1949}].

\begin{figure}[h!]\centering
	\includegraphics[width=0.45\textwidth]{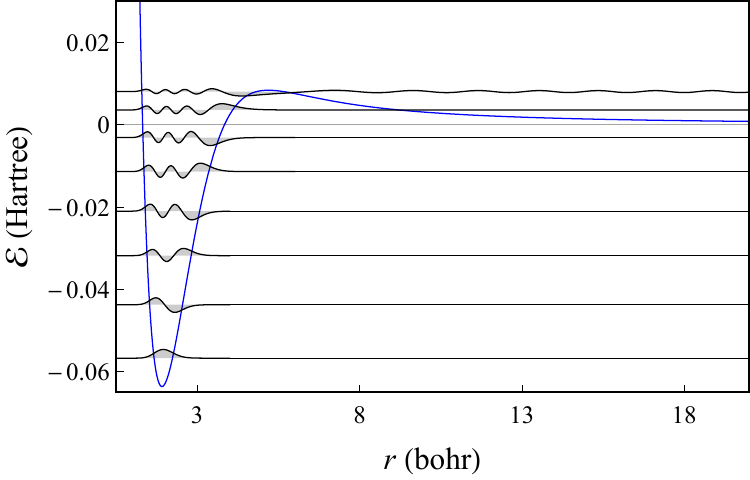}
	\caption{H$_{2}$ effective potential for $l=23$. Bound states (negative energy) and resonance states are shown.}\label{graphics_effective_potential}
\end{figure}

The effective potential, $V_l(r)$, is the sum of a Born-Oppenheimer internuclear potential energy $V(r)$ and a centrifugal term  
\begin{equation}\label{effective_potential}
    V_l(r)=V(r)+\frac{l(l+1)}{2\mu r^2}.
\end{equation}
Typically effective potentials of diatomic molecules grow exponentially as $r$ goes to zero due to internuclear repulsion; for very large values of $r$, the dissociation limit, the effective potential approaches zero from above (below) for $l\ne 0$ ($l=0$). Figure \ref{graphics_effective_potential} shows the effective potential of $\mathrm{H}_2$ molecule for angular momentum $l=23$. In this figure bound states display negative energies, while two resonance states appear with positive energy less than the maximum of the centrifugal barrier.
\\
Equation \eqref{Schrodinger} can be written as a linear system of first order differential equations in the phase-space $\{\psi,\dot\psi\}$:
\begin{subequations}\label{Schrodinger_system}
\begin{align}
    \dot{\psi}(r) &= \varphi(r), \\
    \dot{\varphi}(r) &= -k_l^2\psi(r),
\end{align}
\end{subequations}
with the dotted symbol representing $r$ derivative, $d/dr$, and $k_l=k_l(r,\mathcal{E})$ given by
\begin{equation}\label{kL_equation}
    k_l=\sqrt{2\mu\left(\mathcal{E}-V_l(r)\right)}.
\end{equation}

Equations \eqref{Schrodinger_system} can be expressed in matrix fashion by defining the phase-space vector $\Phi=\left\{\psi,\varphi\right\}$,
\begin{equation}\label{Schrodinger_matrix}
\dot{\Phi}=A_{l}(r,\mathcal{E})\Phi,
\end{equation}
with
\begin{equation}\label{Matrix_A}
A_{l}(r,\mathcal{E})=\begin{pmatrix}
    0 & 1 \\
    -k_l^2 & 0
    \end{pmatrix}.
\end{equation}

Equation \eqref{Schrodinger_matrix} is an initial value problem, its solutions $\Phi=\Phi(r,\Phi_0,\mathcal{E})$, are $r$-parametric phase curves or trajectories. These trajectories are functions of the initial conditions $\Phi_0=\left\{\psi(r_0),\varphi(r_0)\right\}$ and the energy $\mathcal{E}$. Equations \eqref{Schrodinger_system} and \eqref{Schrodinger_matrix} define the tangent vector field $\dot\Phi=\{\dot\psi(r),\dot\varphi(r)\}$. 

\begin{figure}[htp]
\includegraphics[width=0.4\textwidth]{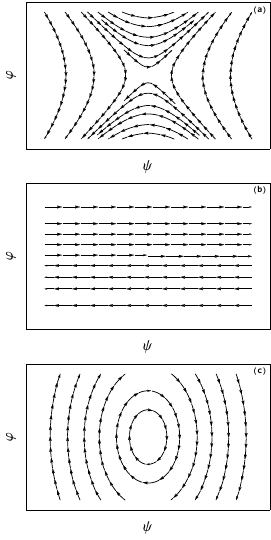}\centering
\caption{Stream plots for trajectories in the vector field $\dot\Phi$. (a) Hyperbolic behavior in classically forbidden regions. (b) Behavior at classical turning points. (c) Elliptic behavior in classically accessible regions.}\label{streamplots}
\end{figure}

The fixed points of system \eqref{Schrodinger_matrix} are given by the solutions, $\Phi^\star$, of $\dot{\Phi}(\Phi^\star)=0$ [\onlinecite{Strogatz1994}]. An inspection of equations \eqref{Schrodinger_system} shows that there is a single fixed point $\Phi^\star$ at the origin,  $\Phi^\star=\left\{0,0\right\}$. The stability of $\Phi^\star$ depends on the eigenvalues of matrix $A_{l}$, \begin{equation}\label{eigenvalues}
\lambda_{1,2}=\mp i k_l.     
\end{equation}
The fixed point $\Phi^\star$ is elliptic for classically accessible regions, $\mathcal{E}>V_l(r)$, or  hyperbolic for classically forbidden regions, $\mathcal{E}<V_l(r)$. At classical turning points, $\mathcal{E}=V_l(r)$, the eigenvalues of $A_l$ are both zero, $\lambda_{1,2}=0$. An inspection of equation \eqref{Schrodinger_system} shows that at the classical turning points, there are lines of non isolated fixed points in phase-space with constant $\varphi$, this can be seen in figure \ref{streamplots}(b). 

The eigenvectors of $A_l$, can be expressed in terms of the eigenvalues $\lambda_{1,2}$, 
\begin{equation}\label{eigenvectors}
  \textbf{e}_{1,2}=\left\{\frac{1}{\lambda_{1,2}}, 1\right\}.  
\end{equation}
For classically forbidden values of $r$, the eigenvectors $\textbf{e}_1$ and $\textbf{e}_2$ define the unstable and stable subspace, respectively. Figure \ref{streamplots}(a) displays the hyperbolic streamlines of the phase-space for a classically forbidden value of $r$. All the hyperbolas of this figure have the same center, located at the origin, and the same asymptotic behaviour. The asymptote with positive slope define the unstable subspace meanwhile, the asymptote with negative slope define the stable subspace. For the effective potential of figure \ref{graphics_effective_potential}, as $r$ goes to zero, the eigenvectors $\textbf{e}_{1,2}$ tend to align with the vertical axis, meanwhile, as $r$ approaches a classical turning point, the eigenvectors $\textbf{e}_{1,2}$ tend to align with the horizontal axis. Figure \ref{streamplots}(b) displays the phase-space structure at the classical turning points. Figure \ref{streamplots}(c) shows the streamlines of the phase-space for classically accessible regions, eigenvalues are complex conjugate pairs, and trajectories follow elliptic orbits around the origin.

\begin{figure}[ht!]\centering
	\includegraphics[width=0.4\textwidth]{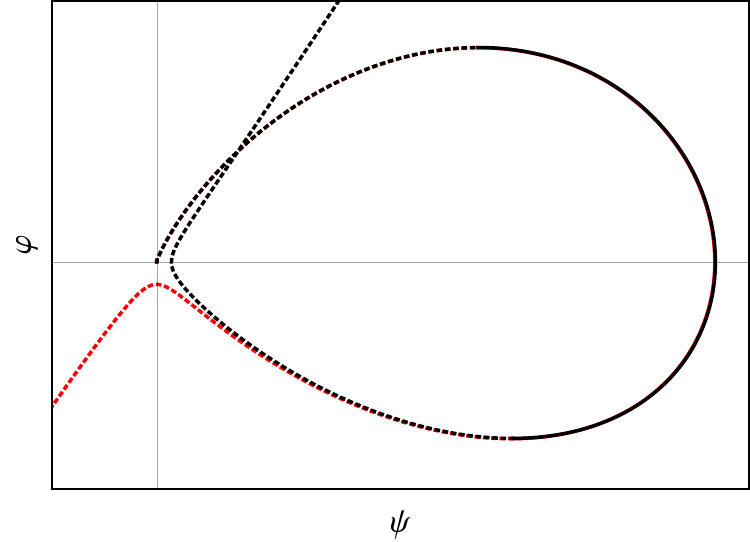}
	\caption{Phase space trajectories for energies around the ground state, $\mathcal{E}_0\mp\Delta\mathcal{E}$ with $l=23$. The black trajectory ($\mathcal{E}_0-\Delta\mathcal{E}$) diverges upwards meanwhile the red trajectory ($\mathcal{E}_0+\Delta\mathcal{E}$) diverges downwards. }\label{phase_space_divergence}
\end{figure}

Equations \eqref{Schrodinger_matrix} are propagated from an initial $r=r_0\approx 0$, located in the classically forbidden region. For a fixed initial condition $\Phi_0=\Phi(r_0)=\{0,1\}$ on the unstable subspace, equations \eqref{Schrodinger_matrix} produce a 1-parameter family of trajectories, $\Phi(\mathcal{E})=\Phi(r,\Phi_0,\mathcal{E})$. The initial condition $\Phi_0$ is obtained from the unstable eigenvector $\mathbf{e}_1=\{\lambda_1^{-1},1\}=\{i/k_l,1\}$. In the classically forbidden region, for $r_0\approx 0$, the unstable vector is approached by $\mathbf{e}_1\approx\{0^+,1\}$. In general, the $\Phi(\mathcal{E})$ are divergent for negative energies, except at the exact energy of the eigenstates. Figure \ref{phase_space_divergence} displays the phase space trajectories $\Phi(\mathcal{E}_0\mp\Delta\mathcal{E})$ corresponding to energies below or above the ground state energy, $\mathcal{E}_0$. The trajectory $\Phi(\mathcal{E}_0-\Delta\mathcal{E})$ diverges upwards meanwhile $\Phi(\mathcal{E}_0+\Delta\mathcal{E})$ diverges downwards. As $\Delta\mathcal{E}$ tends to zero, the divergence of $\Phi(\mathcal{E}_0\mp\Delta\mathcal{E})$ occurs closer to the fixed point $\Phi^\star$. It is interesting to notice how the behavior of trajectories $\Phi(\mathcal{E}\mp\Delta\mathcal{E})$ around eigenenergies resembles the behavior of unstable orbits of dynamical systems [\onlinecite{Strogatz1994}].

\section{Results}

The $X^{1}\Sigma_{g}^{+}$ electronic state of H$_{2}$ molecule is modeled by a Born-Oppenheimer potential energy curve, $V(r)$. This electronic curve is obtained by spline interpolations of three energy data reported for different ranges of the interatomic distance $r$. Sims and Hagstrom [\onlinecite{Sims2006}] in the range $0.4\leq r/a_0\leq6.0$, Wolniewicz [\onlinecite{Wolniewicz1993}] for $6.0< r/a_0\leq10.0$, and Wolniewicz [\onlinecite{Wolniewicz1998}] for $10.0< r/a_0 \leq20.0$. 

The numerical propagation of the equation \eqref{Schrodinger_system} is performed by using the \verb|NDSolve| function of Wolfram Mathematica. The fifth-order Explicit Runge-Kutta method is used with a variable step size ranging from $10^{-6}$ to $10^{-3}$ a.u.

\subsection{Energy-momentum map}

The classical equilibrium solutions, $r_e$, of the effective potential \eqref{effective_potential} are given by the roots of the equation $dV_l(r)/dr=0$ for $l\ge 0$. These roots can be seen as the zero contours of the function $C(l,r)=dV_l(r)/dr$ in the $l-r$ plane. 
\\
\begin{figure}[ht!]\centering
\includegraphics[width=0.45\textwidth]{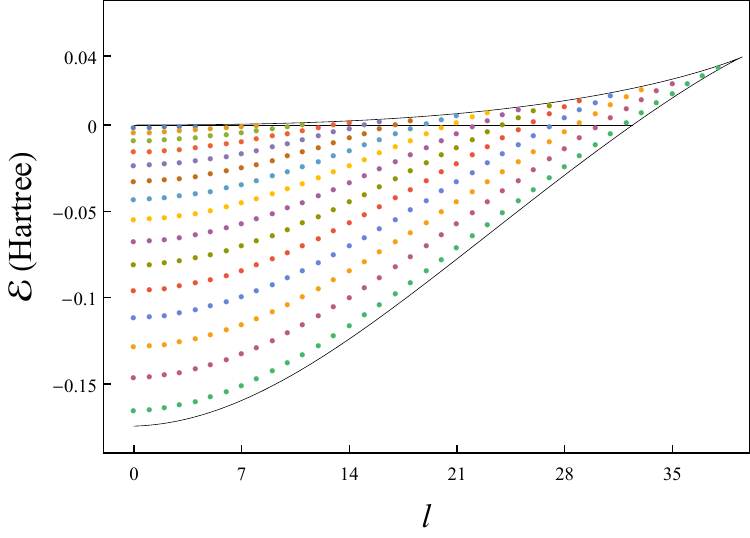}
\caption{Energy momentum diagram ($\mathcal{E}-l$) for H$_2$ molecule with angular momentum $l$. The classical equilibrium curves $V_s(l)$ and $V_u(l)$ are displayed as solid curves. The asymptotic equilibrium, $V_a$, is shown as a straight line with $\mathcal{E}=0$. The lattice of quantum states is displayed with colored points. The color of the quantum states indicates vibrational quantum number.} \label{energy_momentum}
\end{figure}

There are two types of equilibrium solutions for $V_{l\ne0}$: stable equilibrium at the bottom of the potential well, and unstable  equilibrium at the top of the centrifugal barrier. In the dissociative limit there is also a stable equilibrium for $V_{l=0}$ and $V_{l\ne0}$. The effective potential $V_{l=0}$ does not have a centrifugal term and only the stable and the asymptotically stable equilibria are possible.  

The zero contours, $C(l,r)=0$, give the set equilibrium solutions as functions of the angular momentum $r_e(l)=\{r_{s}(l),r_{u}(l)\}$. The effective potential can be evaluated along $r_s(l)$ and $r_u(l)$ producing the curves 
\begin{equation}\label{stable_unstable_curves}
    V_{s,u}(l)=V_l(r_{s,u}(l)),
\end{equation}
on the $\mathcal{E}-l$ plane [\onlinecite{Arango2005}]. Figure \ref{energy_momentum} shows the classical energy momentum equilibria for the effective potential $V_l(r)$. In the figure there are two curves forming a smile pattern [\onlinecite{Arango2004}], and a straight line in black color. The lowest energy curve is clearly the stable equilibrium, $V_s(l)$, meanwhile the highest energy curve is the unstable equilibrium, $V_u(l)$. Between the stable and unstable equilibria there is a straight line with $\mathcal{E}=0$ representing the asymptotic stable equilibrium, $V_a(l)$. As the angular momentum increases the curves $V_s(l)$ and $V_u(l)$ approach each other until they finally join forming the right upper vertex observed in figure \ref{energy_momentum}. This vertex indicates the highest possible value of $l$ that can hold a potential energy well. 

The curves for classical stable and unstable equilibria enclose an area that contains bound and resonance quantum states. The line $V_a(l)$ divides this area in two: the region for bound states with negative energy, and the region for resonance states with positive energy. These regions can be seen in figure \ref{energy_momentum}, the points inside the regions are bound and resonance states, points with the same color have the same vibrational quantum number.

\subsection{Arc length and winding number for finding bound and resonance states}

It is convenient to use a polar representation of the trajectories: $\rho=\left\|\Phi\right\|$, and $\theta= \arctan{(\varphi/\psi)}$. The arc length $\ell(\mathcal{E})=\ell[\Phi(\mathcal{E})]$, and the winding number $w(\mathcal{E})=w[\Phi(\mathcal{E})]$, can be defined in terms of the one-parameter family of trajectories $\Phi(\mathcal{E})$:
\begin{subequations}\label{winding_arclenght}
\begin{align}
 \ell(\mathcal{E}) &=\int_{r_0}^{r_f}{\|\dot{\Phi}(\mathcal{E})\| dr},\\
    w(\mathcal{E}) &=\frac{1}{2\pi}\int_{r_0}^{r_f}{\dot\theta dr}.
\end{align}
\end{subequations}

The use of equations \eqref{Schrodinger_system} in the integrand of $\ell(\mathcal{E})$ gives
\begin{equation}\label{arclength}
    \ell(\mathcal{E})=\int_{r_0}^{r_f}\sqrt{\varphi^2+k_l^4\psi^2}dr.
\end{equation}
The winding number can be expressed in terms of $\psi$ and $\varphi$,
\begin{equation}\label{winding}
    w(\mathcal{E})=-\frac{1}{\pi\mu}\int_{r_0}^{r_f}\frac{h(\psi,\varphi,r)}{\rho^2}dr,
\end{equation}
with $h=h(\psi,\varphi,r)$ as the energy density
\begin{equation}\label{energy_density}
    h(\psi,\varphi,r)=\tfrac{1}{2}\left(\mu\varphi^2+\mu k_l^2\psi^2\right).
\end{equation}

\begin{figure}
\includegraphics[width=0.45\textwidth]{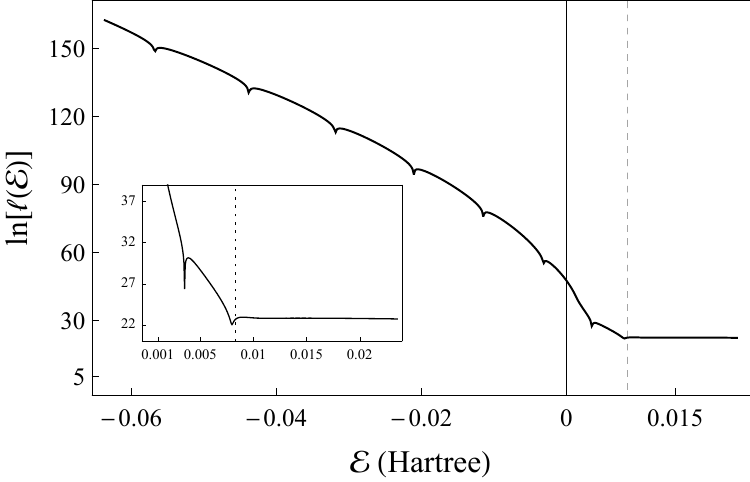}
\caption{Logarithm of the arc length as a function of the energy.  The energy for the top of the centrifugal barrier is displayed as a dashed vertical line. The inset shows the resonances with finer detail.}\label{arclenght_fig}
\end{figure}

Bound and resonance states are phase-space trajectories that minimize the arc length and the derivative of the winding number. Figure \ref{arclenght_fig} displays the arc length, for $\mathrm{H}_2$ with $l=23$, as function of the energy for trajectories with initial condition $\Phi_0=\{0,1\}$. It is evident in this figure that $\ell(\mathcal{E})$ displays downwards spikes at the energies of bound and resonance states. The decrease of $\ell(\mathcal{E})$ for bound states can be explained by analyzing the curves of figure \ref{phase_space_divergence}. Trajectories for bound states approach close loops starting and ending at the fixed point $\Phi^\star$. Trajectories with energy below or above the energy of a bound state diverge exponentially along the unstable subspace, hence the arc length of these trajectories grow exponentially. In the inset of figure \ref{arclenght_fig} is possible to see that the peak of the low energy resonance is sharp and deep, while the high energy resonance is a shallow and wide peak. Although the center of the second resonance is below the top of the centrifugal barrier, the peak of this resonance covers energies below and above the top of the barrier. Figure \ref{arclenght_fig} shows that for energy above the centrifugal barrier the arc length is nearly a constant function of the energy. The decrease in arc length for resonances can be explained using figure \ref{resonances_radial}. The center of a resonances peak is due to the trajectory exiting the centrifugal barrier with the smallest amplitude on an energy interval around the resonance, hence the smallest arc length. Figure \ref{resonances_radial} displays the trajectory at the center of the resonances and trajectories with energy around the center of the resonance.

\begin{figure}[h!]
\includegraphics[width=0.45\textwidth]{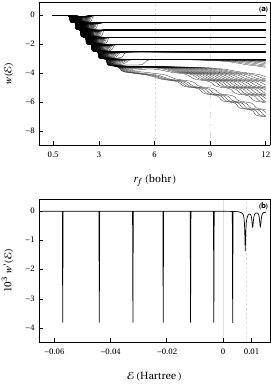}
\caption{(a) Winding number for different energies, as function of the upper integration limit, $r_f$, of equation \eqref{winding}. Vertical dot-dashed and dashed lines indicate the classical turning points for the first and second resonance, respectively. (b) Derivative of the winding number at $r_{f}=6$ bohr for the curves of panel (a). The top of the centrifugal barrier is displayed by a dashed vertical line.}\label{winding_derivative}
\end{figure}

Figure \ref{winding_derivative}a displays the winding number, for $\mathrm{H}_2$ and $l=23$, as function of the upper integration limit,  $r_f$, of equation \eqref{winding}. All the trajectories are propagated from the same initial condition $\Phi_0=\{0,1\}$. In the figure, we can see that for $r_f=6$, the winding number of trajectories with negative energy gather around $w_v\approx v/2$ for $v=0,...,5$; Trajectories with energy below the ground state gather with $w_0\approx 0$, trajectories with energy between the ground state and the first excited state gather with $w_{v=1}\approx0.5$, and so on. Figure \ref{winding_derivative}a shows that trajectories with energy between the highest bound state, $v=5$, and the first resonance, gather transiently with $w\approx -3.05$ at $r_f\approx 6$. Trajectories with energy between the first resonance and the second resonance gather transiently with $w\approx -3.57$ at $r_f\approx 6$. In figure \ref{winding_derivative} it is shown that as the upper integration limit of equation \eqref{winding} grows beyond $r_f>6$, the winding numbers of the trajectories associated to the resonances begin to scatter downwards to smaller values of $w(\mathcal{E})$.

Figure \ref{winding_derivative}b displays the derivative of the winding number, $w'(\mathcal{E})$ at $r_f=6$, for the trajectories displayed in figure \ref{winding_derivative}a. The derivative of the winding number is negative for all the values $\mathcal{E}$, $w'(\mathcal{E})\leq0$, which shows that $w(\mathcal{E})$ is a monotonically decreasing function of the energy. In figure \ref{winding_derivative}b, bound states are shown as peaks with energy $\mathcal{E}_v<0$ for $v=0,...5$. The resonances are the peaks with energies $\mathcal{E}_6=3.52\times10^{-3}$ and $\mathcal{E}_7=7.99\times10^{-3}$ Hartree. The function $w'(\mathcal{E})$ can be used to calculate the lifetime of resonances. Using the full width at half maximum (FWHM) of the resonance peaks we obtained a lifetime of $4.25\times10^{-12}$ s for the first resonance and $1.09\times10^{-13}$ s for the second resonance. The calculation of the resonance lifetimes required an energy discretization $\Delta\mathcal{E}=1.\times10^{-8}$ Hartree. The lifetimes for the same resonances obtained by DVR \cite{Perez2019} are $3.495\times10^{-12}$ and $2.948\times10^{-13}$ s. The two rightmost peaks of figure \ref{winding_derivative}b, with lower intensity, are produced by the set of 5 trajectories with the lowest $w(\mathcal{E})$ in figure \ref{winding_derivative}a. It can be seen in figure \ref{winding_derivative}a that taking larger values of $r_f$ will generate additional peaks in $w'(\mathcal{E})$. Table \ref{tab_resonances} shows energy and lifetime of some resonance states for different values of $l$.

\begin{table}[htb]
\caption{Resonance energies and lifetimes for H$_{2}$ on the $X^1\Sigma_g^+$ electronic state. Columns two and four are the energies and lifetimes obtained in this work. Energies and lifetimes are expressed in Hartrees and seconds respectively.  \label{tab_resonances}}
\begin{ruledtabular}
\begin{tabular}{c c c c c c c}
  $l$   &  Energy & DVR Energy & Lifetime & DVR lifetime \\
   \hline
   4     &  2.759(-6)  &  3.3(-6)    &   9.841(-9)  & 1.137(-10)      \\
   \hline
   17    &  1.040(-3) &  1.041(-3)  &  4.73(-12)  & 6.468(-12)    \\
   \hline
   23    &  3.519(-3)  &  3.520(-3) &   4.25(-12)  & 3.495(-12)     \\
         &  7.989(-3)  &  7.978(-3) &   1.09(-13)  & 2.948(-13)     \\
   \hline
   28    &  4.849(-3)  &  4.849(-3) &  2.77(-12)  & 2.983(-12)   \\
         &  1.121(-2)  &  1.121(-2) &  2.81(-12)  & 1.905(-12)    \\
\end{tabular}
\end{ruledtabular}
\end{table}

\subsection{Phase-space representation of wave functions}

\begin{figure}[htp]
\includegraphics[width=0.45\textwidth]{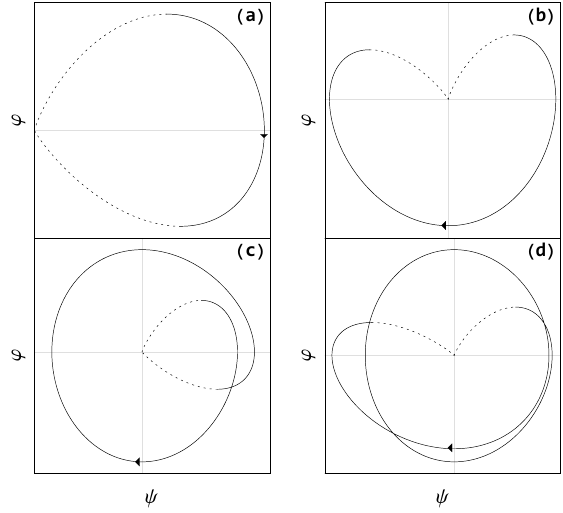}
\caption{Phase space trajectories of H$_2$ with $l=23$ for vibrational eigenstates with (a) $v=0$, (b) $v=1$, (c) $v=2$, and (d) $v=3$. Solid (Dashed) line represent the part of the trajectory propagated on a classically allowed (forbidden) region.}\label{psiresj1314}
\end{figure}

\begin{figure}
\includegraphics[width=0.4\textwidth]{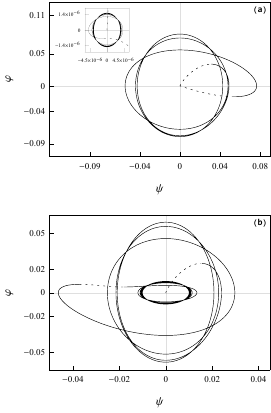}
\caption{Phase space trajectories for the resonance states of H$_2$ with $l=23$. (a) $v=6$, (b) $v=7$. The inset of panel (a) shows the oscillatory behavior for large values of $r$.}\label{respsj23}
\end{figure}

The phase-space trajectories, $\Phi_v=\Phi(\mathcal{E}_v)$, for the lowest four bound states of $\mathrm{H}_2$ at $l=23$ are shown in figure \ref{psiresj1314}. The trajectories $\Phi_v$ are produced by the numerical propagation of equations \eqref{Schrodinger_system} at the energies that minimize $\ell(\mathcal{E})$ and $w'(\mathcal{E})$. All the trajectories of figures \eqref{psiresj1314} and \ref{respsj23} start from the same initial condition,  $\Phi_0\approx\{0^+,1\}$. The first part of the propagation takes place in the  repulsive part of the potential with trajectories following the streamlines o the unstable manifold. The trajectories in the panels of figure \ref{psiresj1314} show this initial part of the propagation as the dashed curve that connects the origin with a point to the right in the upper half of the plane. The propagation continues in the classically accessible region producing the solid curves of figure \ref{psiresj1314}, these curves circulate in clockwise direction. The nodes of the wave functions are the intersection of the trajectories with the $\varphi$ axis, $\psi=0$. Finally there is a transition to the external (right) classical forbidden region, which is represented by the dashed curve approaching to the fixed point $\Phi^\star$ along the stable manifold. 

\begin{figure}
\includegraphics[width=0.4\textwidth]{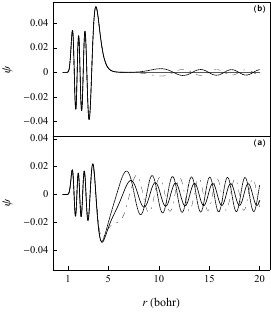}
\caption{Radial wave functions for the $\mathrm{H}_2$ molecule with $l=23$. The resonance, the wave function that minimize $\ell(\mathcal{E})$ and $w'(\mathcal{E})$ are shown in solid line, functions with energy below and above the resonance are shown in dot-dashed and dashed line, respectively. Panel (a) displays the $v=6$ resonance with energy $3.5191\times10^{-3}$ Hartree. Panel (b) shows the $v=7$ resonance with energy $7.9893\times10^{-3}$ Hartree.}\label{resonances_radial}
\end{figure}

The phase space trajectories, $\Phi_6$ and $\Phi_7$, for the two resonance states of $\mathrm{H}_2$ with $l=23$ are shown in figure \ref{respsj23}. The trajectories for resonance states circulate in clockwise direction, as bound states. Figure \ref{respsj23}(a) displays the phase space trajectory for the lower energy resonance, $\Phi_6$. The first resonance displays a strong character of bound state, starting and ending very close to the fixed point $\Phi^\star$; however, as it is shown in the inset of the figure, the trajectory does not approach $\Phi^\star$ but falls in cycle with small amplitude. The trajectory for the second resonance is shown in figure \ref{respsj23}(b). The second resonance start near $\Phi^\star$ and behaves in oscillatory fashion as a bound state, the trajectory then crosses the centrifugal barrier in an almost vertical line and finally falls in a cycle with considerable amplitude. The second resonance displays a mixed character of bound state and plane wave. The wave functions $\psi_v(r)$ for the resonances are shown in figure \ref{resonances_radial}, for the sake of comparison. Panel \ref{resonances_radial}(a) shows the low energy resonance, it is clear from the figure that this resonance resembles a bound state with almost imperceptible oscillations for large values of $r$. The higher energy resonance is shown in figure \ref{resonances_radial}(b) where again is evident its mixed character between a bound state and a plane wave. 

\section{Conclusions}

This phase-space propagation scheme displays advantages over spectral and pseudospectral methods, like DVR or FGH, for calculating bound and resonance states. The stability analysis of the phase-space structure allows to obtain important information that can be related with the features of the wave function in classically allowed or forbidden regions. Bound states behave as classical unstable periodic orbits meanwhile resonance states behave as stable cycles. The use of the arc length and the winding number as functions of the energy, for the same initial condition, produces spectra with peaks centered at the energy of  bound and resonance states. Since phase-space propagation is an initial value method, there is not imposition of artificial boundary conditions or adjustable absorbing potentials. The use of the energy as a control variable facilitates obtaining the derivative of the winding number, $w'(\mathcal{E})$, at high energy resolution. The high-resolution $w'(\mathcal{E})$ allows to obtain accurate lifetimes, and minimal asymptotic amplitude resonances, by direct measure of the FWHM and the minimum of the resonance peaks respectively. The aforementioned advantages of the phase-space propagation method could be helpful for obtaining resonance states of heavy molecules, with centrifugal barriers reaching hundreds of bohrs, as KRb or RbCs. The energy momentum map offers a global view of the quantum states. The curves made by the energy of the classical equilibria as functions of $l$ allow to divide the $\mathcal{E}-l$ plane in regions for bound and resonance quantum states.   
  
\section{Acknowledgments}

This project has been fully financed by the internal research grants of Universidad Icesi. 

\section{Data Availability Statement}

The data that support the findings of this study are available from the corresponding author upon reasonable request.

\bibliographystyle{aipnum4-1}
\bibliography{references}
\end{document}